\documentclass[]{emulateapj}
\usepackage{graphicx,latexsym,natbib,epsfig}

\begin{document}

\title{Swift X-ray Telescope study of the Black Hole Binary MAXI J1659-152: Variability from a two component accretion flow}
\author{M. Kalamkar\altaffilmark{1}, M. van der Klis\altaffilmark{1}, L. Heil \altaffilmark{1}, J. Homan \altaffilmark{2}}
\altaffiltext{1}{Astronomical Institute, ``Anton Pannekoek'', University of Amsterdam, Science Park 904, 1098 XH, Amsterdam, The Netherlands}
\altaffiltext{2}{MIT Kavli Institute for Astrophysics and Space Research, 70 Vassar Street, Cambridge, MA 02139, USA}
\email{maithili@oa-roma.inaf.it}

\begin{abstract}
\noindent We present an energy dependent X-ray variability study of the 2010 outburst of the black hole X-ray binary MAXI J1659-152 with the \textit{Swift} X-ray Telescope (XRT). The broad-band noise components and the quasi periodic oscillations (QPO) observed in the power spectra show a strong and varied energy dependence. Combining \textit{Swift} XRT data with data from the Rossi X-ray Timing Explorer, we report, for the first time, an rms spectrum (fractional rms amplitude as a function of energy) of these components in the 0.5--30 keV energy range. We find that the strength of the low-frequency component ($<$ 0.1 Hz) decreases with energy, contrary to the higher frequency components ($>$ 0.1 Hz) whose strengths increase with energy. In the context of the propagating fluctuations model for X-ray variability,  we suggest that the low-frequency component originates in the accretion disk (which dominates emission below $\sim$ 2 keV) and the higher frequency components are formed in the hot flow (which dominates emission above $\sim$ 2 keV). As the properties of the QPO suggest that it may have a different driving mechanism, we investigate the Lense-Thirring precession of the hot flow as a candidate model. We also report on the QPO coherence evolution for the first time in the energy band below 2 keV. While there are strong indications that the QPO is less coherent at energies below 2 keV than above 2 keV, the coherence increases with intensity similar to what is observed at energies above 2 keV in other black-hole X-ray binaries.
\end{abstract}

\section{Introduction}\label{intro}
\noindent Black hole X-ray binaries (BHB) are systems in which a stellar-mass black hole accretes matter from a companion star. An accretion flow forms around the black hole, along with outflows in the form of collimated jets and disk winds. The accretion flow is believed to have two components: an inner flow/corona (an optically thin medium where photons are Comptonized by hot electrons) and an (optically thick) accretion disk. After decades of studies of the energy spectra and variability of many BHBs, it is generally understood that the interplay between these two components of the accretion flow gives rise to different `states' of the system in an outburst. Phenomenologically, the evolution of the system through these states is understood quite well. We first discuss the behavior of a BHB in outburst in terms of the different phenomena commonly observed and then discuss the existing models developed to explain their origin. We refer the reader to \cite{homan2005}, \cite{remillard2006} and \cite{klis2006} for detailed phenomenology and conventions and to \cite{done2007} for the discussion of models. \\
\\
The states observed during an outburst can be broadly classified as hard and soft states. In the `low' intensity hard state (LHS), the energy spectrum is dominated by hard emission from the hot flow (a term we use to refer to the corona/the inner flow/base of the jet, without preference for any model), and the power spectrum is characterized by strong broad band noise (fractional \textit{rms} amplitude up to $\sim$ 50 \%). The disk emission and the intensity increase when the source makes a transition to the intermediate state (IMS), which can be divided into hard and soft IMS (HIMS and SIMS, respectively). The energy spectrum is softer in the SIMS compared to the HIMS, while the fractional \textit{rms} amplitude is stronger in the HIMS (up to $\sim$ 30 \%) compared to the SIMS (few \%). During the HIMS the type-C Quasi Periodic Oscillations (QPOs, peaked narrow components) are detected in the power spectrum, while the SIMS is often accompanied by one of the two different types of QPOs, type-A or type-B QPOs (see \citealt{wijnands1999qpos}; \citealt{remillard2002qpos}; \citealt{casella2005}, for QPO classification).  Multiple transitions between the IMSs are often seen in BHBs before the source goes into the high soft state (HSS). In the HSS, the X-ray spectrum is dominated by soft disk emission and  variability is very weak. At some point in time the intensity decreases  and eventually the source goes back to the LHS through the IMS. It should be noted that not all sources show all these states.\\
\\
Although there is a reasonably clear picture of the phenomenological behavior, some major and important  physical aspects of the accretion flow are not fully understood. There is no agreement about the structure and origin of the hot flow, or on the disk geometry \citep{done2007}. While there is progress in modelling (see below), our understanding of the origin of variability remains incomplete. Most of the variability studies in the past decade were performed with the Rossi X-ray Timing Explorer (\textit{RXTE}) mission. The Proportional Counter Array (PCA) on board \textit{RXTE} covered the energy range 2-60 keV. The hot flow emission dominates this energy band in the hard state during which the strongest variability is observed. Based on the hard band variability studies, many models have been proposed to explain its origin, which we discuss below. \\
\\
The propagating fluctuations model \citep{lyub1997} proposes that fluctuations of mass accretion rate modulate the X-ray emission, giving rise to the observed variability. These fluctuations can arise and propagate throughout the flow and modulate the X-ray emission produced in the inner regions. \cite{chu2001} showed that as the fluctuations propagate to smaller radii on local viscous time scales, high frequency fluctuations are suppressed due to viscous damping. This means that low frequency fluctuations generated at large radii can propagate to smaller radii and modulate the emission. Higher frequency fluctuations can only survive if generated at smaller radii. As the emission from the inner regions dominates at higher energies, the amplitude of high frequency variability is therefore stronger at high energies than at low energies \citep{kotov2001}. Further works \citep[see e.g.,][and the references therein]{ingram-lt-2011} associated different frequencies of the broad band noise in the hard state power spectrum with different radii; the lower break frequency is associated with the outer radius of the hot flow (truncation radius of the disk) and the upper break frequency (which we will refer to as the hump) is deeper in the hot flow. They also associate the frequency of the type-C QPO with the Lense Thirring precession of the hot flow \citep{stella1998, fragile2007}. \\
\\
In the works discussed above, variability is attributed to the hot flow and the disk is considered unimportant for variability studies. However, recently \cite{wilkinson} and \cite{kalamkar1753.5} using \textit{XMM-Newton} and \textit{Swift}, respectively (which can access energies down to 0.3 keV) showed that the disk contributes significantly to variability at energies $<$ 2 keV on time scales longer than a few seconds. They suggested that the propagating fluctuations could arise intrinsic to the disk giving rise to variable emission in the soft band. This highlights the importance of access to the soft band for variability studies. The power spectra have been observed to be dramatically different (at energies above 2 keV) along the various states of the outburst. A difference in the behavior of the power spectra below 2 keV in different outburst states was shown for the BHB SWIFT J1753.5--0127 \citep{kalamkar1753.5} with \textit{Swift}. However, this source is peculiar as it does not show a typical outburst progressing through different states. Similar investigations for sources which show more typical outbursts are necessary, before the application of the above models can be generalized to all BHBs. \\
\\
In this manuscript, we report energy dependent variability studies of the outburst of the BHB MAXI J1659--152 with \textit{Swift} observations that cover the 0.5-10 keV energy range. Section \ref{1659literature} introduces the source and discusses earlier reports. The \textit{Swift} data used for this study along with the \textit{RXTE} results from \cite{kalamkar1659} are discussed in Section \ref{obdata}. In Section \ref{results}, we present the results of the  variability analysis, the evolution and correlations of different power spectral components in two sub-bands of the X-ray Telescope (XRT): 0.5-2 keV and 2-10 keV, along with \textit{RXTE} PCA results in the 2-60 keV band from \cite{kalamkar1659}. We present our interpretation and discuss the origin of variability in the context of the models in Section \ref{origin-var}, followed by our conclusions and summary of results in Section \ref{discussion}. \\
%
\subsection{Earlier reports on MAXI J1659--152}\label{1659literature}
\noindent MAXI J1659-152 (henceforth J1659) was discovered on 2010 September 25 with the \textit{Swift} Burst Alert Telescope \citep[BAT;][]{barthelmy1659} and identified as a new Galactic X-ray transient (\citealt{ugarte1659atel}; \citealt{negoro1659}). It was soon identified as a stellar-mass black hole candidate as it exhibited a type-C QPO in  \textit{RXTE} observations  \citep{kalamkar1659atel}. \cite{kuulkers1659} determined an orbital period of 2.41 hr, making J1659 the shortest known orbital period BHB. The accretion disk inclination is estimated to be 60-80 degrees and the companion star is suggested to be an M5 dwarf star. Various reports estimate a distance in the range of 5-8.6 kpc and a height of 2.4 kpc above the Galactic plane (\citealt{kuulkers1659}; \citealt{kennea1659}; \citealt{yamaoka1659}). The evolution of the source along the hardness-intensity diagram and its variability properties showed that J1659 behaved  similar to other BHBs \citep{kalamkar1659, teo16592011}.   \\
\\
\cite{wenfei1659} report energy dependent variability studies with \textit{Swift} and \textit{RXTE}, similar to our analysis. The main difference in our works is that we report the full evolution of \textit{all} power spectral components in individual observations (or GTIs), their correlations, and energy dependence along the outburst. In addition to the discussion of the origin of the broad band variability, where we both arrive at similar conclusions (see Section \ref{discussion}), we also discuss the origin of the QPO.  \\

\begin{figure}
\center
\includegraphics[width=4.8cm,height=7.2cm,angle=-90]{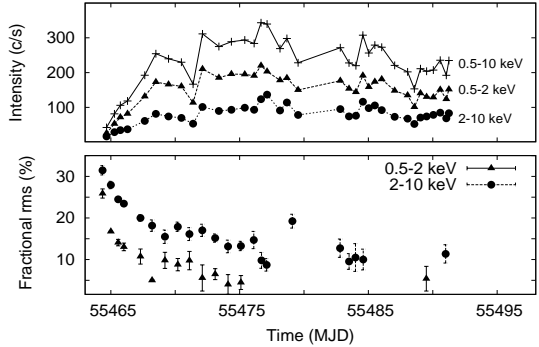}
\caption{Top panel - Light curve in the full energy band and two sub-bands as indicated; Bottom panel -  Evolution of the  fractional rms amplitude integrated up to 10 Hz in the energy bands indicated. Each point in the light curve represents one observation and is pile-up, bad pixel and background corrected. See Section \ref{obdata} for the details of the evaluation of the fractional rms amplitude.\\}
\label{lc}
\end{figure}

\section{Observations and data analysis}\label{obdata}
\noindent We analyzed all 38 observations taken in Windowed Timing (WT) mode with the X-Ray Telescope \citep[XRT;][]{burrows2005} on board the \textit{Swift} satellite between September 25, 2010 (MJD 55464) and October 22, 2010 (MJD 55491). Observations lasted between 0.9 and 19.5 ks containing between 1 and 28 Good Time Intervals (GTIs) of 0.1-2.5 ks. The data were obtained in the WT mode data (in wt2 configuration), which has a time resolution of 1.766 ms. We processed the raw data  using the standard procedure discussed in \cite{evans2007} and selected only grade 0 events. Pile-up, bad pixel corrections and background corrections were applied to the light curves. For comparison, we also use the results from the first 47 \textit{RXTE} \citep{jahoda2006} observations taken between September 28, 2010 (MJD 55467) and October 22, 2010 (MJD 55491), the same period as the XRT observations, in the 2--60 keV energy band as previously presented in \cite{kalamkar1659}.  \\
\\
\noindent To generate the XRT power spectra, we determine the source region and remove the data that is at the risk of pile-up. This is done by removing the central pixel, and if necessary, additional pairs of pixels symmetrically around the central pixel, until the count rate is below 150 c/s \citep[see][for more details]{kalamkar1753.5}. Leahy-normalized \citep{leahy1983} fast Fourier-transform power spectra were generated using 115.74-s continuous intervals (no background or bad pixel corrections were applied). The 1.766 ms time resolution gives a Nyquist frequency of 283.126 Hz.  As the first four observations consist of multiple long individual GTIs (some a few hundred seconds long) we report their individual power spectra. For the rest of the observations, we report the average power spectrum per observation. To facilitate comparison with \textit{RXTE} which covers 2--60 keV (henceforth xte band), two energy bands were used: hard, 2--10 keV (also covered by \textit{RXTE}) and soft, 0.5--2 keV (not covered by \textit{RXTE}). See \cite{kalamkar1659} for the details of \textit{RXTE} power spectrum generation. Periods of dipping activity in the X-ray light curve, reported by \cite{kuulkers1659short} were not excluded from our analysis.\\
\begin{figure}
\center
\includegraphics[width=6.1cm]{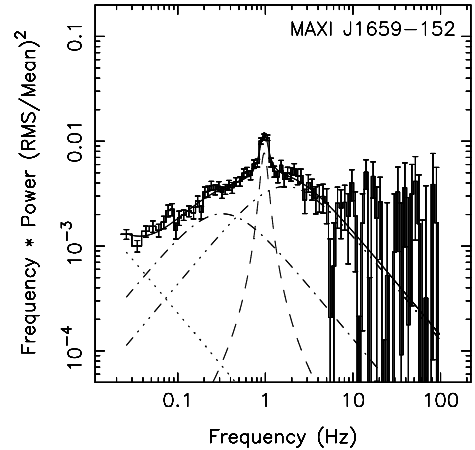}
\caption{Representative power spectrum (obs-id 00434928003, MJD 55466) in the 0.5-10 keV band. The best fit model using multiple Lorentzians is shown. The components, in the order of increasing frequency, are indentified as low-frequency noise, break, type-C QPO and hump. The power spectrum has been rebinned for presentation.}\label{pds}
\end{figure}

\noindent A drop-off in power above 100 Hz due to instrumental effects has been reported in the XRT power spectra \citep{kalamkar1753.5}. We also observe this drop-off in our data on this source. Hence, we analysed the power spectra in the frequency range $<$100 Hz only. As the Poisson level deviates from the expected value of 2.0, we estimate the Poisson level by fitting a constant between 50-100 Hz where no source variability is observed \citep{kalamkar1753.5}. This estimated Poisson level is subtracted and the power spectra are expressed in rms normalization \citep{Vanderklis89}. The power spectra are fitted with  several Lorentzians in the ``$\nu_{max}$" representation \citep*{belloni2002}. We fit for the following parameters: the characteristic frequency $\nu _{\rm max} \equiv \nu _{0}\sqrt{1+1/(4Q^{2})}$, the quality factor Q$\equiv\nu_{0}$/FWHM, and the integrated power $P$, where $\nu_{0}$ is the centroid frequency and FWHM is the full width at half maximum of the Lorentzian. When Q turned out negative, it was fixed to 0 (i.e., we fitted a zero-centred Lorentzian); this did not significantly affect the other parameters. We only report components with a single-trial significance $P/\sigma_{P_-}$ $>$ 3.0 (unless otherwise stated),  with $\sigma_{P_-}$ the negative error on $P$ calculated using $\Delta\chi^2$ = 1. All the errors reported in this work, including the Figures, are 1$\sigma$ errors.
\section{Results}\label{results}
\subsection{Light curve and variability evolution}\label{lc-var}
\noindent Figure \ref{lc} (top panel) shows the light curve in the 0.5-10 keV energy band and the two sub-bands: soft and hard. The light curve has been reported to be of the fast-rise exponential decay type in \textit{Swift} BAT \citep{kennea1659} and \textit{RXTE} PCA observations \citep{yamaoka1659}. As Swift began observing the source $\sim$ three days before \textit{RXTE}, we can report the early rise of the outburst. We observe that the source was already in the HIMS during the first XRT observation, as there was strong broad band noise (up to 30 \% fractional rms amplitude) and a type-C QPO \citep[see also][]{kalamkar1659}. The peak intensity was observed on MJD 55476.7. Transitions to the SIMS (where a type-B QPO is detected) were observed twice with \textit{RXTE} \citep{kalamkar1659}; the first excursion to the SIMS on MJD 55481.7 was not observed by \textit{Swift}, the second transition on MJD 55484.7 was covered by XRT observations but these ended before the transition back to the HIMS on MJD 55501. \cite{kalamkar1659} reported that the source did not make a transition to the HSS  \citep[state where spectrum has thermal disk contribution above 75\% and the variability is weak;][]{remillard2006} before returning to the hard state while \cite{teo16592011} report the transition of the source to the HSS \citep[softest spectrum dominated by thermal disk component and with weak variability;][]{belloni-2010-lnp}. \\ 
\\
The evolution of the fractional rms amplitude (henceforth referred to as \textit{rms})  integrated up to 10 Hz in the soft and the hard bands is shown in the bottom panel of Figure \ref{lc}. It is consistent with the integrated \textit{rms} reported in \cite{kennea1659}. The integrated \textit{rms} was 31.5 $\pm$ 1.1\% in the hard band during the first observation, consistent with what is expected in the HIMS, and 26 $\pm$ 1.1 \% in the soft band.  It decays in both energy bands as the source evolves towards the SIMS. The two excursions to the SIMS reported by \textit{RXTE} were accompanied by a drop in the integrated \textit{rms} in the xte band \citep[see Figure 1 in][] {kalamkar1659}. The first excursion on MJD 55481.7 was not covered by \textit{Swift}, but after the second transition at MJD 55484.7, the integrated \textit{rms} was 10.0 $\pm$ 2.5 \% in the hard band (7.1 $\pm$ 2.7 \% in the xte band). During the rest of the observations the \textit{rms} stayed close to $\sim$ 10 \% in the hard band (between 3\%-9\% in the xte band till MJD 55491).  The soft band variability is poorly constrained from MJD 55476.1 - MJD 55489 and hence not reported here. It should be noted that the integrated \textit{rms} is higher in the hard band than in the soft band for all XRT observations. 

\begin{figure}
\includegraphics[width=8cm,height=9cm,angle=-90]{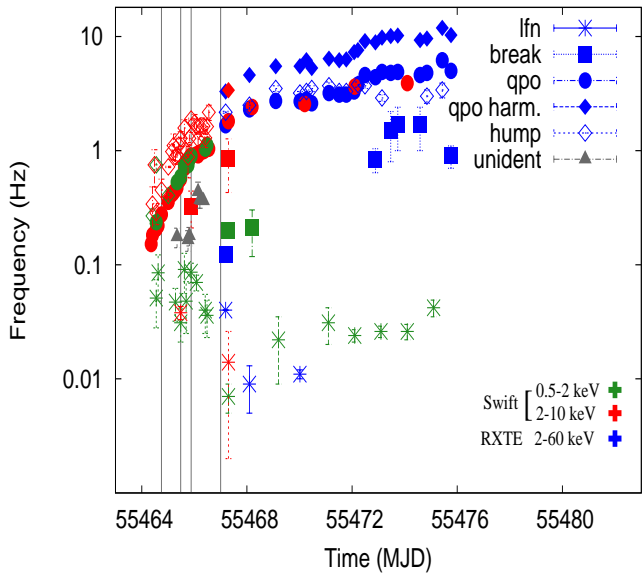}
\caption{The frequency evolution of all power spectral components with time. The grey lines indicate the end-time of first four XRT observations in which we report detections in individual GTIs. The rest of the detections are per observation. Components are indicated by different symbols. The different colours indicate the energy bands: 0.5-2 keV (soft), 2-10 keV (hard) and 2-60 keV (xte). The unidentified components are detections in the soft band.}\label{freq-time}
\end{figure}

\subsection{Power spectral evolution}\label{pds-evolution}
\noindent Figure \ref{pds} shows a representative power spectrum of an XRT observation in the 0.5-10 keV energy band. The different components, in the order of increasing frequency, can be identified as: the low frequency noise (lfn), the `break' component, the QPO identified as the type-C QPO (which will be referred to as the QPO), and the broad band noise (referred to as `hump') underlying the QPO. The harmonic of the QPO is also detected (not present in the power spectrum shown here). The power spectrum  is very similar to the ones exhibited by other BHBs in the HIMS  (e.g., \citealt{homan2005}; \citealt{casella2005}). The coherences Q are in the range of 0.0--1.2 for the lfn, 0.1--0.2 for the break, 0.4--11.7 for the QPO and 0.0--1.93 (and one incidence of a high Q at 5.7) for the hump. All the components are detected in the hard and the soft bands, although not always simultaneously and not in every observation (see Table \ref{table:para}). A type-B QPO has been reported in the xte band with \textit{RXTE} \citep{kalamkar1659}, but we do not detect it in the XRT power spectra. 
\begin{figure*}
\includegraphics[width=6cm,height=7.25cm,angle=-90]{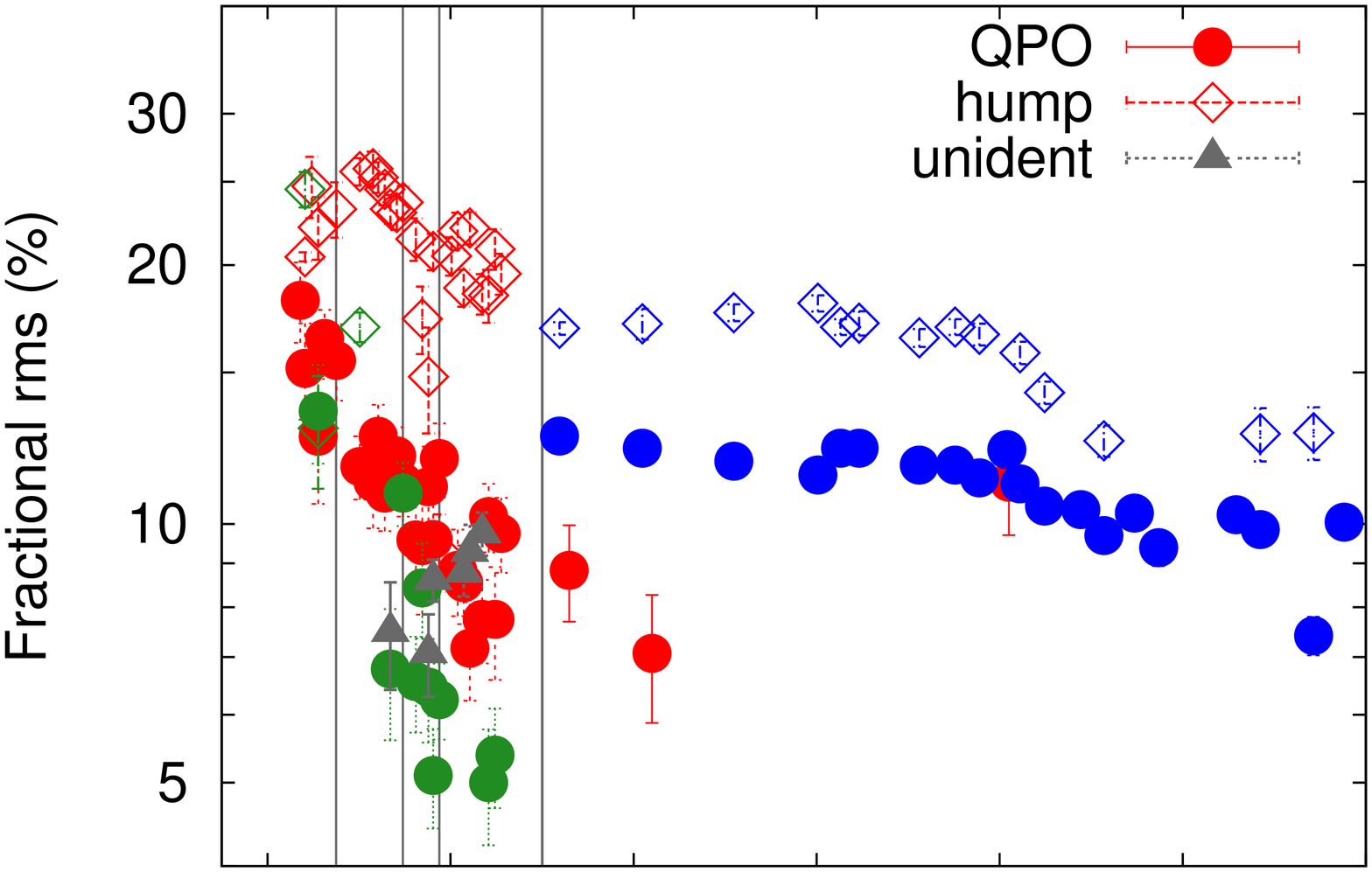}\includegraphics[width=6cm,height=5.5cm,angle=-90]{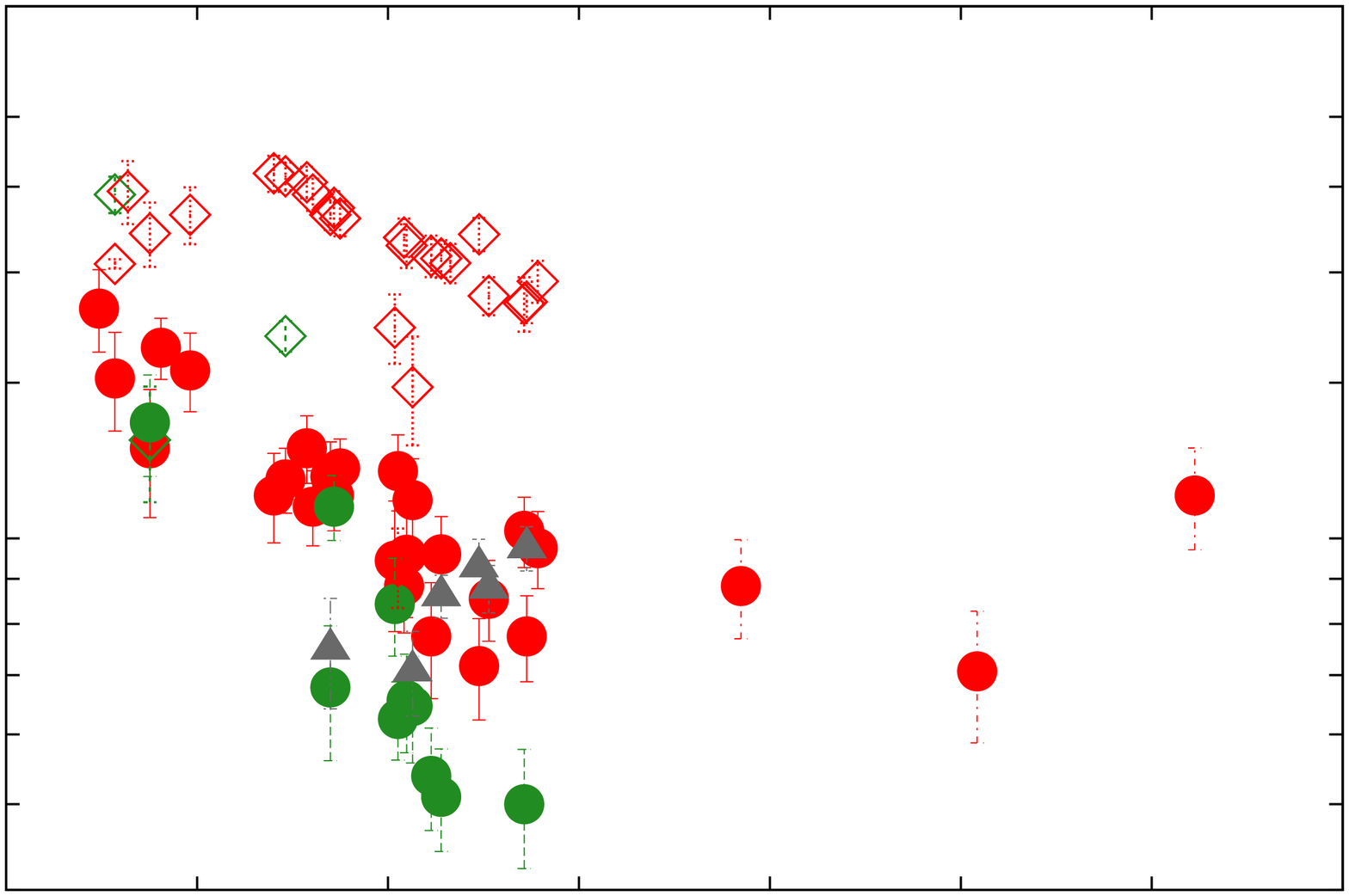}\includegraphics[width=6cm,height=5.5cm,angle=-90]{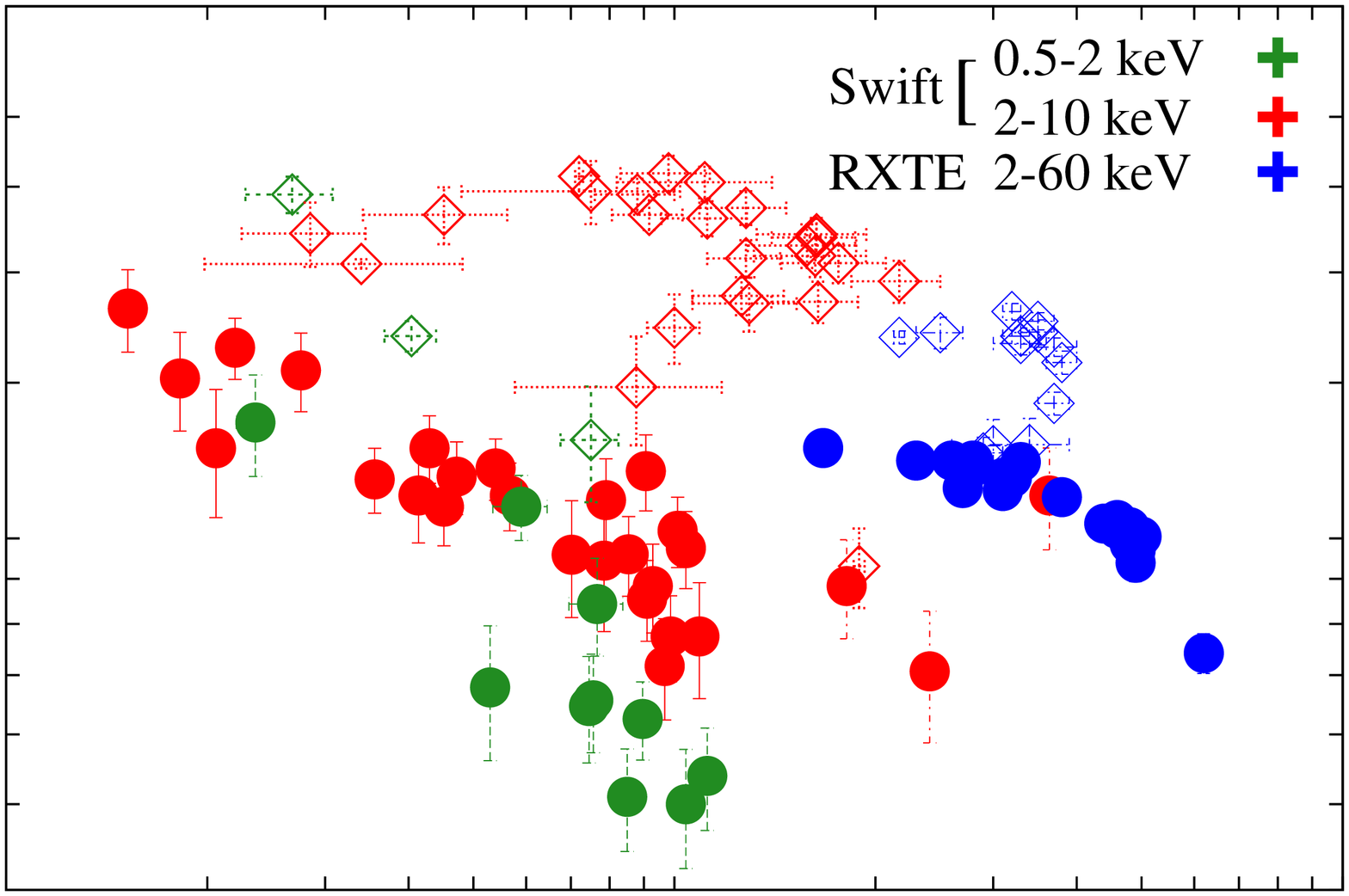} 
\includegraphics[width=6cm,height=7.25cm,angle=-90]{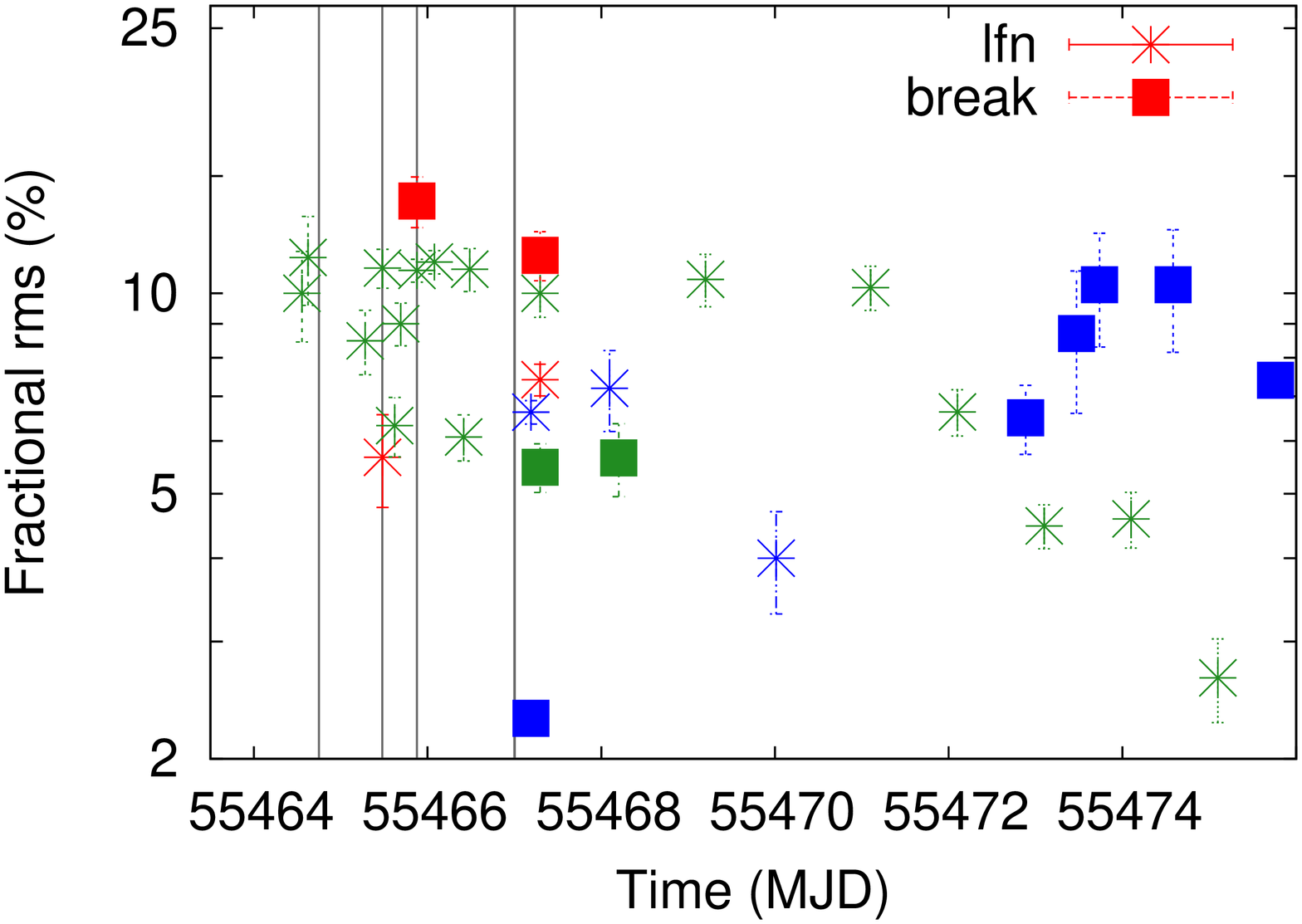}\includegraphics[width=6cm,height=5.5cm,angle=-90]{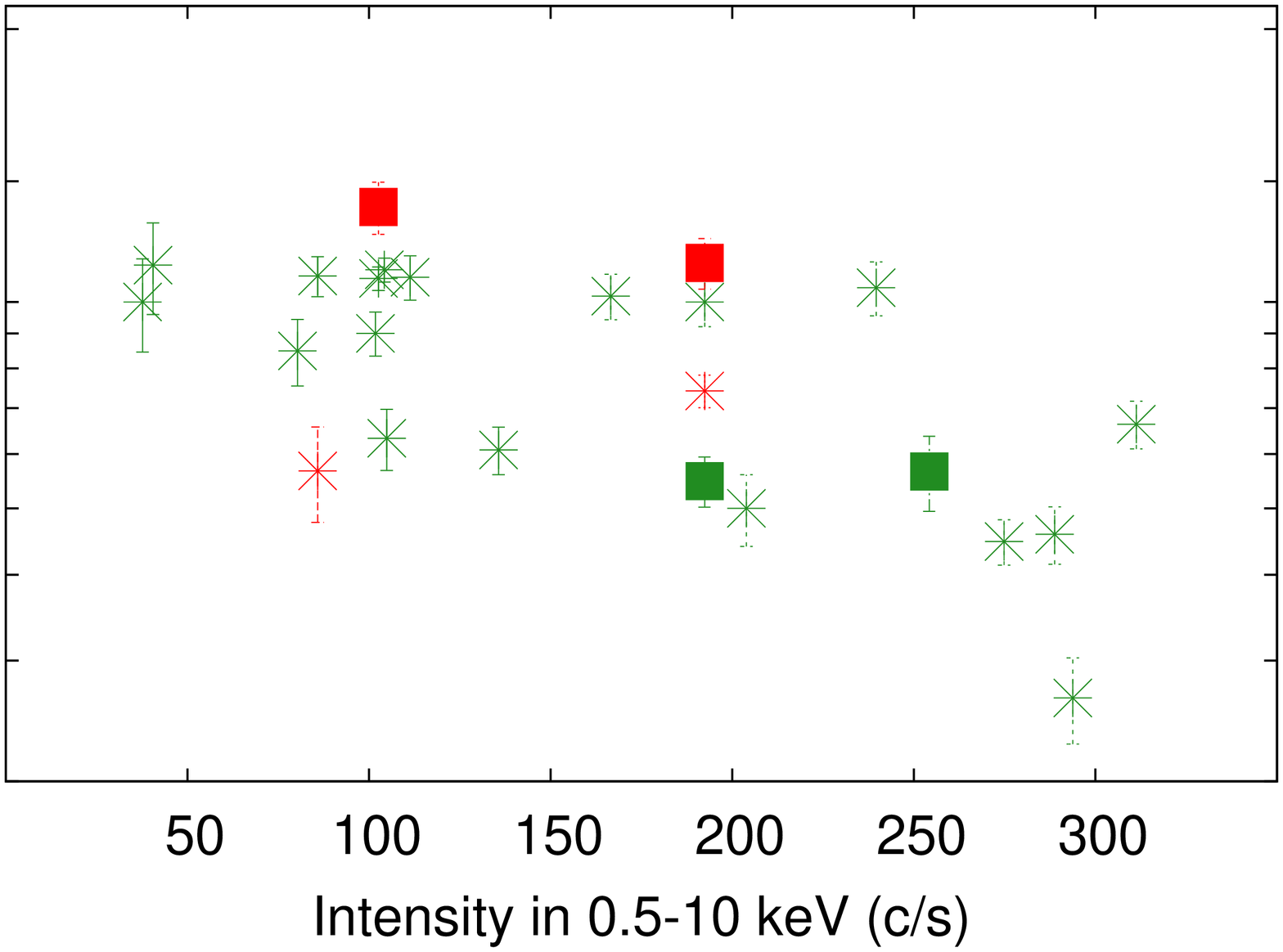}\includegraphics[width=6cm,height=5.5cm,angle=-90]{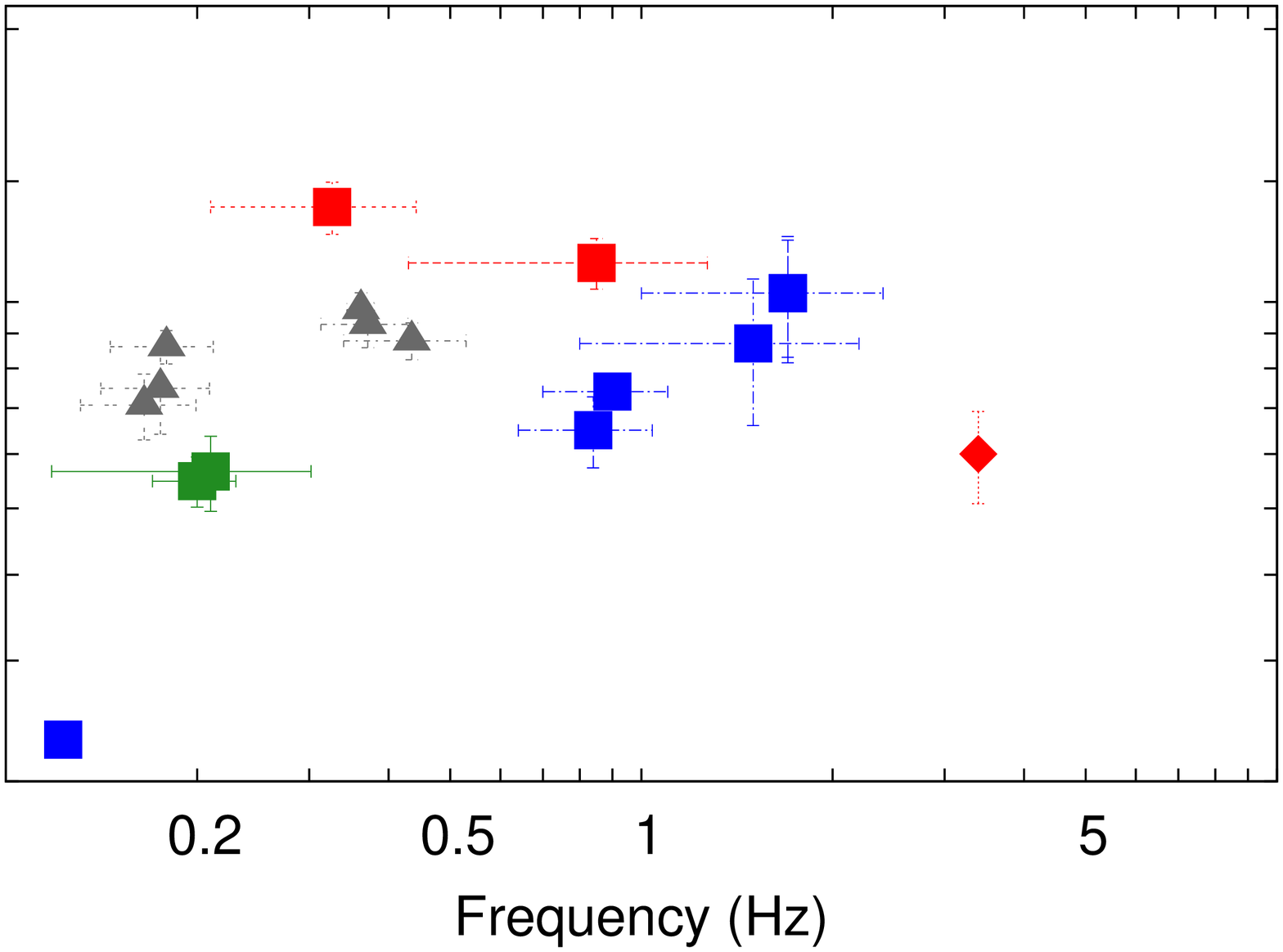} 
\caption{Evolution of fractional rms amplitude of the different components with time (left panels), and its dependence on XRT intensity (middle panels) and frequency of the respective components (right panel). In the left panels, the grey lines indicate the end-time of first four XRT observations in which we report detections in individual GTIs. The rest of the detections are per observation. The components are as indicated in the left panels with colors indicating the energy bands as shown in the top right panel.  The different components are plotted in different panels for the purpose of clarity.}\label{rms-evol}
\end{figure*}

\subsection{Evolution of the parameters and their energy dependent behavior}\label{param-evol}
\subsubsection{Frequency evolution with time}\label{freq-evolution}
\noindent The evolution of the frequencies of all the components in the soft and the hard bands with time, along with the xte band from \cite{kalamkar1659} is shown in Figure \ref{freq-time}. The vertical grey lines mark the (end-time of the) first four observations for which we report detections in the individual GTI. The rest of the detections are in each average \textit{Swift} and \textit{RXTE} observation. For clarity, additional noise components detected only in the xte band reported in \cite{kalamkar1659} have been omitted in this Figure. \\
\\
All the components, except the lfn, show an increase in frequency as the outburst progresses. The QPO frequency evolution in the hard and soft band is consistent with the reports of \cite{kennea1659} and \cite{wenfei1659} with \textit{Swift} XRT.  The rise in the QPO and hump frequency is very rapid during the first three days. The hump component is detected in the hard and soft band till MJD 55465 and 55466.6 respectively, and its frequency is higher than the QPO frequency in all detections. In the xte band, the hump frequency is higher than the QPO frequency up to MJD 55472, after which it is lower than the QPO frequency till its last detection at MJD 55575.4. The frequency of the QPO and the hump in both the soft and the hard bands show correlations with intensity (not shown here), which also increases with time; this behavior is commonly seen in BHBs \citep[see, e.g.,][]{klis2006}.   \\
\\
The break component has very few detections. It shows an increase in frequency in the hard and xte band. A change in frequency is not clearly seen between the two detections in the soft band. The components shown in grey are detections in the soft band that cannot be identified unambiguously; these could be the break, hump or lfn. Lack of simultaneous detections of all components in the soft band power spectra makes it difficult to identify them correctly in some of the observations. \\
\\
The lfn component does not follow the same evolution as the rest of the components. This component is seen consistently at low frequencies below 0.1 Hz across changes in intensity and state transitions without a large increase (of more than a decade) in frequency like the rest of the components. There are more detections of this component in the soft band compared to the hard band.  It was also detected in the soft band at 0.050 $\pm$0.008 Hz at MJD 55489.5, when the source was in the SIMS (not shown in Figure \ref{freq-time}). In the xte band, it can be fit with a Lorentzian only in three observations. In some of the other observations, the power at low frequencies can be constrained by a power law with a slope varying between 1.96 - 3.4 (with \textit{rms} randomly varying in the range of $\sim$ 2-4 \%).
\subsubsection{Rms evolution with time}\label{sec-rms-time-evol}
\noindent The \textit{rms} evolution of the different components with time is shown in the left panels of Figure \ref{rms-evol}. The \textit{rms} shows a general trend of decrease in strength with time for most of the components in all three bands. The individual components show the following behavior:\\
\\
$\bullet$ The hump is the strongest amongst all the components. Unlike other components, it first shows an increase in strength till MJD 55465.2, followed by a decay. This behavior closely follows the 15--150 keV BAT light curve that shows a sharp rise reaching the peak at MJD 55465, which is  much earlier than the XRT peak, followed by a decay \citep{kennea1659}. In the soft band, the hump does not follow the BAT light curve and shows a decay similar to the rest of the components. \\
$\bullet$ The \textit{rms} of the QPO in the hard and soft band shows a decay in amplitude with time, which is steeper than that of the hump. During the first \textit{RXTE} observation quasi-simultaneous with XRT, the QPO was stronger in the xte band than the hard band, with no detection in the soft band. Overall, the QPO is weakest and decays most rapidly in the softer bands.\\
$\bullet$  The break component shows an \textit{rms} decrease in the hard band, not much change in the soft band, and an initial increase in strength followed by a decrease in the xte band. So, the break component becomes stronger at higher energies, but much later in the outburst. \\
$\bullet$  The lfn shows a decrease in \textit{rms}, but the fall is not monotonic, particularly in the soft band. Its \textit{rms} is higher in the soft band than the hard band for the simultaneous detections. In the xte band when the power at low frequencies is constrained by a power law, the \textit{rms} randomly varies in the range of 2-4 \% and is always lower than the \textit{rms} in the hard and the soft bands.

\subsubsection{Rms evolution with intensity}\label{sec-rms-int-evol}
\noindent The middle panels in Figure \ref{rms-evol}  show the \textit{rms} dependence of all components in the soft and the hard band on XRT intensity in the 0.5-10 keV band. All the components (except hump in the hard band) show an anti-correlation with intensity. The \textit{rms} of the hump in the hard band first shows a rise and then a decay, associated with its non-monotonic behavior versus time in the rise of the outburst. This results in a weaker correlation with intensity in the 0.5-10 keV band than of the other components. The lfn shows a decrease, but with a large scatter, indicating that the dependence on intensity is very weak. 

\subsubsection{Rms evolution with frequency}\label{sec-rms-freq-evol}
Figure \ref{rms-evol} (right panels) shows the relation between the \textit{rms} of different components and their corresponding frequencies. As the lfn does not show strong evolution in frequency, it is omitted here. The break component in the xte band is the only component for which the \textit{rms} shows a positive correlation with frequency for all detections. In the hard band, the break appears to have an anti-correlation.  The behavior of the break component in the soft band is unconstrained by our data. 
The \textit{rms} of the QPO shows an anti-correlation with its frequency in all three bands. Interestingly in the xte band, the anti-correlation becomes steeper when the \textit{rms} falls below $\sim$10 \%. This happens close to the time around which the hump frequency falls below the QPO frequency (Figure \ref{freq-time}, MJD 55472) seen in the xte band; there are no hump detections in the soft and the hard band during this period.\\
\\
We refer to the QPO frequency discussed above as the turnover frequency. The hump component behaves differently below and above this turnover frequency - below the turnover frequency, the \textit{rms} of the hump does not show dependence on its frequency, i.e., the relation is flat, while above the turnover frequency the \textit{rms} and the hump frequency decrease in a correlated fashion in the xte band. In the hard band for the hump, although the shape of the track is somewhat reminiscent of that in the xte band, it should be noted that these observations were taken a few days before the xte band ones (see the left panels). Also, in the hard band the value of the \textit{rms} jumps between the `flat' and `correlated' branches several times; the behaviour is not chronological as is the case in the xte band. In the soft band the hump shows a linear anti-correlation. If the unidentified detections in the soft band are the hump, then the track will have a similar two branch shape traced chronologically like the xte band but earlier in time. This degenerate behavior of the hump  \textit{rms} versus frequency and the shape of the tracks followed in different energy bands has not been reported before for this source. 
\\
\subsubsection{Coherence of the QPO}\label{qpo}
\begin{figure}
\includegraphics[width=5.5cm,height=9cm,angle=-90]{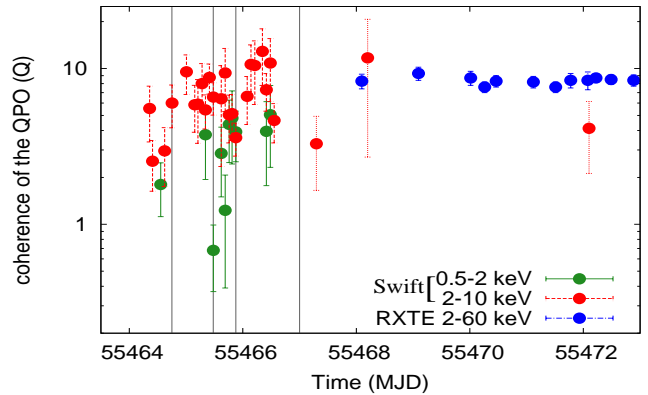}
\caption{Evolution of the coherence Q of the QPO with time in three energy bands as indicated. The grey lines indicate the end-time of first four XRT observations in which we report detections in individual GTIs. The rest of the detections are per observation.} \label{q-time}
\end{figure}

\noindent Figure \ref{q-time} shows the evolution of the coherence Q of the QPO in all three bands.  We report for the first time  the evolution of the Q in the soft band. It increases during the rise of the outburst, similar to the hard band. There has been no evidence of QPO frequency dependence on the energy \citep{belloni1997} but the fast rise in the QPO frequency may lead to the broadening of the component, resulting in a lower than intrinsic Q. In our data, the rate of change of frequency during the first three days is 5.4 $\times 10^{-6}$ Hz/s, in the hard as well as the soft bands. For a typical GTI $\sim$ 1 ks long, contribution to the broadening of the QPO due to increase in frequency is 0.0054 Hz (maximum of 0.0135 Hz for the longest GTI of 2.5 ks). The total FWHM of the QPO are in the range of 0.13--0.87 Hz in the soft band and 0.027--0.25 Hz in the hard band. Hence, the increase in frequency contributes to the broadening of the QPO, but by a small factor in most cases and importantly, by the same magnitude in the hard and soft bands. It is interesting to note that the Q in the soft band is lower than in the hard band for all simultaneous detections, sometimes significantly so. The weighted mean Q value of only the simultaneous detections in the hard and soft band are 5.07$\pm$0.29 and 1.23$\pm$0.07, respectively. Also, as shown earlier, the QPO is weaker in the soft band than in the hard band. This suggests that the QPO is broader as well as weaker in the soft band than in the hard band.

\subsection{The rms spectrum and energy dependence of frequency}\label{rms-spec}
\begin{figure}
\includegraphics[width=7.5cm,height=10cm,angle=-90]{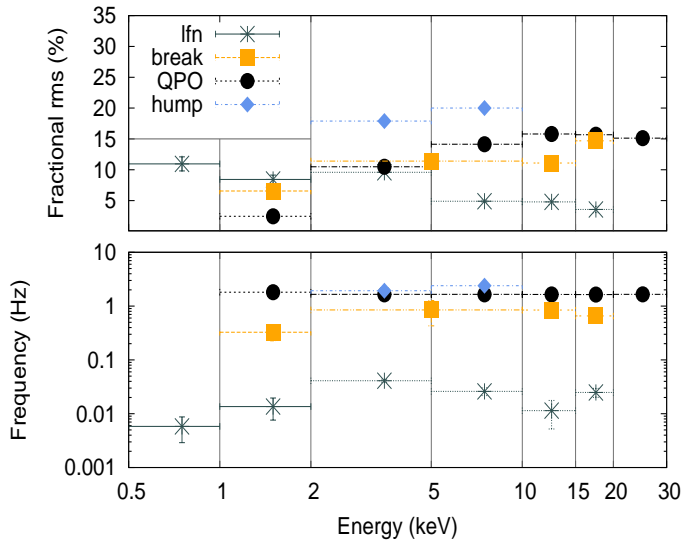} 
\caption{The dependence of the fractional \textit{rms} amplitude (top panel) and the corresponding frequency (bottom panel) of each component on energy from the first (quasi) simultaneous XRT and \textit{RXTE} observation (MJD 55467). The vertical grey lines indicate the boundaries of the energy bands and the points are plotted at the central energy bin. The detections below 2 keV are from the XRT data and the detections above 2 keV are from the \textit{RXTE} data, except for the break shown in the 2-10 keV range which is from the XRT data.}\label{rms-spectrum}
\end{figure}

\noindent Figure \ref{rms-spectrum} shows the dependence of the \textit{rms} and frequency of the corresponding components on energy for the first (quasi) simultaneous observation with XRT and \textit{RXTE}. This is the first report of the \textit{rms} spectrum, i.e., \textit{rms} as a function of energy, in the 0.5--30 keV energy range. We generate power spectra in various energy bands shown in Figure \ref{rms-spectrum}. We fit the power spectrum in each energy band with multiple Lorentzians; when Q turned out negative it was fixed to 0.0, which is the case for the lfn. We then calculate the \textit{rms} of each component and plot it as a function of the corresponding energy band. When the \textit{rms} of a component cannot be constrained in an energy band, we exclude that point from these plots. In the \textit{rms} spectrum (top panel), the lfn is strongest in the 0.5--1 keV band, where no other component is detected. The component is significantly detected till 20 keV with a decreasing \textit{rms}; in the 20-30 keV band, the integrated \textit{rms} up to 0.1 Hz is 2.8 \%. Hence, the lfn has a soft spectrum.\\
\\
The rest of the components show the opposite behavior; their \textit{rms} increases with energy. The hump is the strongest component but is detected only in the 2-10 keV bands (although it shows strong indications of being harder, see Section \ref{sec-rms-time-evol}). The break component is detected in  the 1-20 keV bands, and its amplitude increases with energy. The QPO, which is the only narrow component in the power spectrum, shows an \textit{rms} increasing with energy till 15 keV and then shows (possibly) a small decrease till 30 keV.  \cite{shaposh1659} report the \textit{rms} spectrum using the same \textit{RXTE} observation. Our results are consistent with a hard spectrum they observe for the QPO. The soft \textit{rms} spectrum of the lfn and hard \textit{rms} spectrum of other higher frequency components is similar to that seen in SWIFT J1753.5--0127 \citep{kalamkar1753.5} up to 10 keV in the XRT data, and like in that source, suggests these components have different origin (Section \ref{origin-var}).\\

\begin{table}
  \caption{Parameters of the rms spectrum shown in Figure \ref{rms-spectrum}. The frequency and fractional rms amplitude of the lfn, the break, the QPO and the hump respectively are shown in various energy bands.}  \label{table:para}
  \begin{center}
    \leavevmode
    \begin{tabular}{lll} \hline \hline 
Energy (keV)  & Frequency (Hz) & Frac. rms ampl. (\%)\\ \hline
& 0.006$\pm$0.003 & 10.95$\pm$1.14 \\
0.5 - 1 &\hspace{0.8cm}- &\hspace{0.8cm}- \\
& \hspace{0.8cm}- &\hspace{0.8cm}- \\
& \hspace{0.8cm}- &\hspace{0.8cm}- \\
\hline
& 0.014$\pm$0.006 & 8.43$\pm$0.71 \\ 
 1 - 2 & 0.33$\pm$0.10 & 6.56$\pm$0.45\\
& 1.81$\pm$0.20 & 2.45$\pm$0.61 \\
 & \hspace{0.8cm}- & \hspace{0.8cm}-\\
 \hline
& 0.041$\pm$0.003 & 9.59$\pm$0.15 \\
2 -5 & 0.85$\pm$0.42 & 11.40$\pm$0.97\\
& 1.651$\pm$0.004 & 10.49$\pm$0.21 \\
& 1.93$\pm$0.07 & 17.89$\pm$0.28 \\
\hline
& 0.026$\pm$0.004 & 4.90$\pm$0.16\\
5 - 10 & \hspace{0.8cm}- &\hspace{0.8cm} -\\
& 1.650$\pm$0.003 & 14.14$\pm$0.17 \\
& 2.39$\pm$0.05 & 20.00$\pm$0.23 \\
\hline
& 0.011$\pm$0.006 &4.80$\pm$0.46 \\
10 - 15 & 0.84$\pm$0.15 & 11.09$\pm$1.15\\
& 1.643$\pm$0.004 & 15.81$\pm$0.25 \\
&\hspace{0.8cm}- & \hspace{0.8cm}-\\
\hline
& 0.025$\pm$0.005 & 3.57$\pm$0.52\\
15 - 20 & 0.656$\pm$0.090 & 14.70$\pm$0.68 \\
& 1.641$\pm$0.006 & 15.68$\pm$0.3\\
& \hspace{0.8cm}- & \hspace{0.8cm}-\\
\hline
& \hspace{0.8cm}- & \hspace{0.8cm}-\\
20 - 30 &\hspace{0.8cm}- & \hspace{0.8cm}-\\
& 1.650$\pm$0.012 & 15.13$\pm$0.56 \\
& \hspace{0.8cm}- & \hspace{0.8cm}-\\ \hline
    \end{tabular}
\end{center}
\end{table}

Figure \ref{rms-spectrum}, bottom panel, shows the energy dependence of the (characteristic) frequency of the components discussed above. It is interesting to note that the QPO, the only narrow component, is the only component whose frequency does not show dependence on energy. Lack of frequency dependence on energy ($>$ 2 keV) for the QPO was also reported by \cite{belloni1997} in GS 1124-68 and GX 339-4. The frequency of the rest of the components show a possible energy dependence.  The hump frequency does not show strong energy dependence with only two detections (both have Q of 0.08). The break frequency (Q in the range of 0-0.18) shows an increase with energy till 15 keV, followed by a possible decrease. A similar energy dependence ($>$ 2 keV) of the break was reported earlier by \cite{belloni1997} in GX 339-4 and GS 1124-68 and in XTE J1650-500 \citep{kalemci2003-1650}. The lfn frequency shows a possible energy dependence. It increases with energy till 5 keV, and appears to decrease at higher energies, however it cannot be said conclusively due to large errors.\\

\section{Origin of variability}\label{origin-var}
The different variability components can be broadly separated into two categories: a) components that evolve in frequency - the QPO, the hump and the break, referred to as the higher frequency components and, b) the component which does not evolve much in frequency - lfn which stays below 0.1 Hz. The \textit{rms} spectrum (Figure \ref{rms-spectrum}) of these two categories also shows different behavior; the higher frequency components are harder, i.e., amplitudes increase with energy, while the lfn is soft, i.e., the amplitude decreases with energy. This suggests that the lfn and the higher frequency components arise in different regions of the accretion flow and/or have different driving mechanisms. We investigate this in the context of the propagating fluctuations model \citep{lyub1997}, and the hot flow Lense-Thirring precession model  \citep{fragile2007, ingram-lt-2011}.  
\subsection{Origin of the low frequency noise}\label{dis-lfn}
\noindent In our analysis, we find that the lfn does not show strong evolution in frequency with either time or intensity. We observe that the lfn \textit{rms} is strongest in the 0.5-1 keV band and decreases with energy. Generally variability is associated with the inner regions of the hot flow/corona. If the lfn originated in the hot flow, then the \textit{rms} would be expected to a) increase with energy similar to higher frequency components and, b) be weaker in the 0.5-1 keV band due to  contamination from non modulated photons from the disk. We see the exactly opposite energy dependence and propose that this component originates in the disk. It was also suggested by \cite{wenfei1659} that the lfn (which they refer to as the power law noise) originates in the disk in this source. \\
\\
The lfn shows all the characteristics of a component originating due to mass accretion rate fluctuations arising in the thermal disk \citep{lyub1997, wilkinson}. The lack of frequency dependence on intensity can be naturally explained as the lfn is not associated with a `moving' radius in the accretion flow. As the source evolves towards the soft state, the inner radius of the accretion disk is suggested to decrease \citep{kennea1659}, but if the fluctuations arise further out in the accretion disk than the truncation radius, the frequency may stay stable. As the fluctuations can propagate to inner regions of the accretion flow, the detection of this component at hard energies (up to 30 keV) can be naturally explained. The drop in \textit{rms} along the outburst could be due to the fluctuations becoming inherently weaker as the source evolves to softer states, or dilution due to stronger unmodulated disk emission, or a combination of both factors. It is not understood why and in what capacity, these factors play a role in decreasing the strength of this component in the soft state. 
\subsection{Origin of higher frequency components}
\subsubsection{Origin of the broad components}\label{dis-hump}
\noindent The set of higher frequency components consists of the type-C QPO, the hump and the break component. Similar to other BHBs, these components are detected in the HIMS. We discuss here the behavior of the broad components viz. the hump and the break. The break component has very few detections in all three bands. So any interpretation should be taken with caution. The break frequency increases in the hard and xte band, but not in the soft band. The break frequency has been associated with the truncation radius of the disk \citep{ingram-lt-2011}. As stated earlier, evolution towards the soft state is thought to be associated with the motion of the accretion disk towards the black hole leading to a decrease in the inner radius. As the disk radius decreases, the frequency of the break increases. The frequency also shows energy dependence. In Figure \ref{freq-time}, the first simultaneous detection (MJD 55467) in all three bands is at different frequencies. It has a higher frequency in the hard band than the soft band (as also seen in the \textit{rms} spectrum), but in the xte band the frequency is the lowest. This could be a fitting artefact as possibly the hump, which is not detected with XRT in this observation, subsumes it. We speculate this as there are more detections of the break in the xte band later when the \textit{rms} of the hump is low, and also the break shows an increase in \textit{rms} over that period. The increase in peak frequency with energy (Figure \ref{rms-spectrum}) can be attributed to the dependence of the emission profile of the energy spectrum on the radius of the accretion disk. \\
\\
Extending further the scenario of propagating fluctuations to smaller radii and more inner regions of the accretion flow, we expect to observe higher frequency variability which is harder in nature \citep{kotov2001, wilkinson, ingram-lt-2011}. The hump is the strongest component in the hard and xte bands (see Figure \ref{rms-evol}), where the emission from the hot flow dominates. Its \textit{rms} follows the BAT light curve in the 15-150 keV band more closely than the XRT light curve in the 0.5-10 keV band. The frequency of this broad component is higher than the break at all times and the QPO for most of the detections (also see below).  All this suggests the origin of the hump to be in the hot flow. This has also been suggested by \cite{wenfei1659}. The origin of fast variability was associated with hard emission by \cite{teo16592011} from their hard band variability studies. In SWIFT J1753.5--0127, the hump was suggested to arise in the hot flow based on its hard \textit{rms} spectrum in the 0.5-10 keV band  \citep{kalamkar1753.5}.   \\
\\
The frequency \textit{rms} correlation of the hump shows a degenerate behavior; the \textit{rms} which is initially at similar values for a range of frequencies, eventually starts decreasing as the frequency decreases. Similar frequency \textit{rms} correlations have been studied in many BHBs \citep{pott-cygx-12003, axelsson2006, klein2008}. They suggest that the emitting region can act as a `filter' to high frequency fluctuations (\citealt{psaltis2000}) and reduce their amplitude. The dampening effects can play a significant role in shaping the power spectrum \citep{kotov2001}. As the source evolves towards softer states, the frequencies of most of the components in the power spectrum increase, moving through this `frequency' filter. The suppression of variability at high frequencies could effectively lead to what appears to be a `lower' peak frequency in our fits. This may explain the behavior in the xte band. This behavior however, cannot explain what we see in the hard band, as the path traced in this correlation is not chornological. It should be noted that this behavior, in some of the works mentioned here, have associated this effect with state transitions, while for J1659, we observe the turnover during the HIMS.
\subsubsection{Origin of the QPO}
\noindent The QPO (and its harmonic) is the only narrow component observed in the power spectrum. Similar to other BHBs, the QPO shows an increase in frequency and a decrease in the \textit{rms} as the source evolves towards soft states. The decrease in \textit{rms} is steepest in the soft band. There are more detections in the hard band than the soft band. It has a hard \textit{rms} spectrum as well. The model of propagating fluctuations naturally predicts the origin of the broad components, but as noted by \cite{ingram2009}, an additional mechanism would be required to explain the high coherence of (only) the QPO. Also, it cannot explain why the QPO frequency does not show energy dependence while the other broad components do. All this indicates that a different mechanism is at play in generating the QPO.\\
\\ 
The Lense-Thirring precession of the hot inner flow \citep{stella1998, fragile2007, ingram2009} is a strong candidate model to explain the origin of QPO (see \citealt{straaten2003}, \citealt{altami-lt-ns-2012} for arguments against the applicability of this model to some neutron star systems). The physical model (\citealt{ingram-lt-2011}) that was developed for the QPO can explain some properties such as the frequency and coherence evolution. High QPO amplitudes at energies higher than 2 keV have been reported earlier in the \textit{rms} spectra of many BHB (see e.g.,  \citealt{belloni1997}; \citealt{sobol-qpo-spectra2006}) which can also be explained by this model. We extend the \textit{rms} spectrum down to 0.5 keV, where we cannot constrain the QPO below 1 keV. A drop in the amplitudes at low energies due to dilution from disk emission was predicted by \cite{ingram2012-frame-dragging}. This was also reported in SWIFT J1753.5--0127 \citep{kalamkar1753.5}. There are strong indications for J1659 that the QPO is narrower in the hard band compared to the soft band. There is no explanation in the model yet for a lower coherence in the soft band.\\

\section{Summary and Conclusions}\label{discussion}
\noindent This work highlights the importance of BHB variability studies with Swift XRT. With \textit{Swift} XRT observations of the black hole binary MAXI J1659-152 during its outburst in 2010,  we report the evolution of all variability components observed in the soft (0.5-2 keV) band simultaneously with the hard (2-10 keV) band. We also present a comparison with the \textit{RXTE} results in the 2-60 keV band from \cite{kalamkar1659}. The merit of this study is that variability is studied over the full energy range 0.5-60 keV. This range contains emission from both components of the accretion flow: the accretion disk which generally dominates below $\sim$ 2 keV and the hot flow which generally dominates above $\sim$ 2 keV. We summarise our results below: \\
\\
1. The integrated broad band noise (0.009-10 Hz) in the hard band is stronger than the soft band for all observations.\\
2. The frequency of the lfn component, which has more detections in the soft band than the hard band, does not show evolution with time and intensity; the \textit{rms} shows a decrease but with a large scatter.\\
3. The break, the hump and the QPO frequency increase with time and intensity; the \textit{rms} decreases with intensity with different behavior for each component, with steeper fall in the soft band compared to the hard band.\\
4. We report for the first time  on the coherence of the type-C QPO down to 0.5 keV and find evidence for lower Q at low energies.\\
5. We present for the first time the \textit{rms} spectrum of different variability components in the 0.5-30 keV energy range. The strength of the lfn decreases with energy, while the strength of the break, the QPO and the hump increases with energy. The frequency of the broad band noise components varies with energy, while that of the type-C QPO is independent of energy. \\
\\
We have strong indications for variability arising in both the components of the accretion flow. In addition, there is also evidence for two different mechanisms at play to generate variability. We investigate our results in the context of propagating fluctuations model for the broad band noise. We suggest that the lfn originates in the accretion disk, while the break and the hump components arise in the hot flow. Many properties of the QPO can be understood in the context of the Lense-Thirring precession of the hot flow model. Hence, \textit{energy dependent} variability study is a powerful method to probe the dynamics of the accretion flow. Other techniques such as lag studies \citep[see e.g.,][where they explore frequencies above 0.1 Hz]{teo16592011} should also be exploited. Extending these studies to lower energies and lower frequencies  are necessary to confirm our results. Such studies and observations of more BHBs with \textit{Swift} can help resolve the long standing question of origin of variability. \\
\\
Study of variability in the soft band can also be useful in probing changes (if any) in the geometry of the accretion flow during state transitions. Dramatic changes in the power spectrum are observed during state transitions, which happen close to the radio flaring behavior episodes, although a causal connection has not been established \citep{fender2009}. A drop in the fractional rms amplitude of broad band variability is observed during state transitions. The radio flaring is associated with the discrete ejections of material, possibly the corona (see e.g., \citealt{rodrig2008corona}; \citealt{fender2009}). One way to probe this is to trace the soft band variability during these ejection events: the variable  disk emission should remain observable during the state transitions, if the ejected material is the corona and does not affect the disk. As part of the variable emission comes from the disk, the drop in variability in the soft band should be less than in the hard band. For J1659, close to the radio flaring behavior, \citep[MJD 55477,][]{vanhorst1659atel}, the hard band variability is detected around 10 \% during that period, but the soft band variability is poorly constrained  (MJD 55476.1 - MJD 55489) and hence we cannot comment on this. However, we would like to remark that monitoring of the soft X-ray variability during radio flaring behavior of BHBs can provide constraints on the ejection of corona scenario. 

\section*{Acknowledgements}
\noindent This research has made use of data obtained from the High Energy Astrophysics Science Archive Research Center (HEASARC), provided by NASA's Goddard Space Flight Center, and also made use of NASA's Astrophysics Data System. This work made use of data supplied by the UK Swift Science Data Centre at the University of Leicester.


\begin{table}
  \caption{The time of the observation, the \textit{Swift} Observation ID, the frequency and fractional rms amplitude in the 0.5-2 keV and 2-10 keV bands of the lfn, the break (only when present and is indicated with a $*$),the QPO and the hump components, respectively. The first four observations show the parameters in multiple GTIs. Only significant ($>$ 3 $\sigma$) detections are reported here.  }  \label{table:para}
    \leavevmode
    \begin{tabular}{llllll} \hline \hline 
Time (MJD)  & Obs.ID & \hspace{1.5cm}0.5-2 keV & & \hspace{1.5cm}2-10 keV \\ 
  & & Frequency (Hz) & rms (\%) & Frequency (Hz) & rms (\%)\\ \hline \hline
55464.359 & 00434928000 & - & - & - &  -\\
& & - & - & 0.152$\pm$0.004 & 18.19$\pm$1.95\\
& & - & - & - & -\\ \hline
55464.409 &  & - & -& - &  -\\
& & - & - & 0.182$\pm$0.008 & 15.17$\pm$1.95\\
& & 0.268$\pm$0.040 & 24.50$\pm$1.16 & 0.340$\pm$0.142 & 20.45$\pm$0.25 \\ \hline
55464.483 &  & - &  - &  - & - \\
& & - & - & - & -\\
& & - & - & 0.750$\pm$0.270 & 24.70$\pm$2.04\\ \hline
55464.553 &  & 0.051$\pm$0.023 & 10.00$\pm$1.5 &  - &  -\\ 
& & 0.236$\pm$0.015 & 13.53$\pm$1.78 & 0.206$\pm$0.006 & 12.65$\pm$2.10\\
& & 0.750$\pm$0.075 & 12.92$\pm$1.94 & 0.285$\pm$0.060 & 22.14$\pm$1.85\\ \hline
55464.622 &  & 0.085$\pm$0.037 & 11.31$\pm$1.72 & - &  - \\
& & - & - & 0.220$\pm$0.008 & 16.43$\pm$1.31\\
& & - & - & - & -\\ \hline
55464.754 &  & - & -& - &  -  \\
& & - & - & 0.276$\pm$0.006 & 15.49$\pm$1.58\\
& & - & - & 0.452$\pm$0.110 & 23.24$\pm$1.72\\ \hline
55465.008 & 00434928001 & - & -&  - &  -\\
& & - & - & 0.356$\pm$0.003 & 11.66$\pm$0.99\\
& & 0.404$\pm$0.036 & 16.94$\pm$0.67 &  0.72$\pm$0.01 & 25.69$\pm$0.93\\ \hline
55465.153 &  & - & -&  - &  - \\
& & - & - & 0.414$\pm$0.010 & 11.18$\pm$1.30\\ 
& & - & - & 0.98$\pm$0.15 & 25.88$\pm$1.22 \\ \hline
55465.209 &  & - & -&  - &  - \\
& & - & - & 0.430$\pm$0.005 & 12.65$\pm$1.11\\
& & - & - & 1.110$\pm$0.290 & 25.30$\pm$1.03\\ \hline
55465.282 &  & 0.047$\pm$0.015 & 8.49$\pm$0.94 & - &  - \\
& & - & - & 0.452$\pm$0.006 & 10.86$\pm$1.06\\
& & - & - & 0.88$\pm$0.12 & 24.50$\pm$1.02\\ \hline
55465.343 &  &- & -&  - &  - \\
& & 0.530$\pm$0.018 & 6.78$\pm$1.18 & 0.472$\pm$0.008 & 11.75$\pm$1.11\\
& & - & - & 0.92$\pm$0.11 & 23.24$\pm$0.90\\ \hline
55465.412 &  & - & -&  - &  -\\
& & - & - & 0.54$\pm$0.006 & 12.00$\pm$0.96\\
& & - & - & 1.12$\pm$0.16 & 23.02$\pm$1.04\\ \hline
55465.481$^*$ &  & 0.031$\pm$0.010 & 10.91$\pm$0.73 & 0.038$\pm$0.005 & 5.67$\pm$0.090\\
& & - & -& 0.038$\pm$0.008 & 5.67$\pm$0.90 \\
& & 0.590$\pm$0.055 & 10.86$\pm$0.92 & 0.567$\pm$0.009 & 11.18$\pm$0.98\\
& & - & - & 1.28$\pm$0.19 & 23.66$\pm$1.04\\ \hline
55465.621 & 00434928002 & 0.091$\pm$0.035 & 6.33$\pm$0.65 &  - &  -\\
& & 0.756$\pm$0.035 & 6.56$\pm$0.84 & 0.702$\pm$0.020 & 9.58$\pm$1.45\\
& & - & - & 1.58$\pm$0.29 & 21.45$\pm$1.21\\ \hline
55465.690 &  &  0.048$\pm$0.023 &9.00$\pm$0.67 & - &  -\\
& & 0.766$\pm$0.071 & 8.43$\pm$1.07&  0.785$\pm$0.012 & 9.43$\pm$1.59\\
& & - & - & 1.000$\pm$1.090 & 17.32$\pm$1.56\\ \hline
55465.759 &  & - & -&  - &  -\\
& & 0.745$\pm$0.020 & 6.46$\pm$0.89 & 0.790$\pm$0.018 & 11.05$\pm$1.26\\
& & - & - & 0.877$\pm$0.300 & 14.83$\pm$2.09\\ \hline
55465.811 &  & - & -&  - &  -\\
& & 0.850$\pm$0.026 & 5.10$\pm$0.68&  0.854$\pm$0.017 & 9.59$\pm$0.99\\
& & - & - & 1.28$\pm$0.16 & 20.74$\pm$1.01\\ \hline
55465.877 &  & 0.086$\pm$0.017 & 10.82$\pm$0.43 &  - &  -\\
& & 0.896$\pm$0.028 & 6.25$\pm$0.63 & 0.906$\pm$0.020 & 11.92$\pm$1.18\\
& & - & - & 1.890$\pm$0.036 & 9.30$\pm$0.96\\ \hline
55466.010 & 00434928003 & - & -&  - &  -\\
& & - & - & - & -\\
& & - & - & 1.759$\pm$0.313 & 20.45$\pm$1.05\\ \hline
    \end{tabular}
\end{table}
\begin{table*}
    \leavevmode
    \begin{tabular}{llllll} \hline \hline 
Time (MJD)  & Obs.ID & \hspace{1.5cm}0.5-2 keV & & \hspace{1.5cm}2-10 keV \\ 
  & & Frequency (Hz) & rms (\%) & Frequency (Hz) & rms (\%)\\ \hline \hline
  55466.078 &  & 0.070$\pm$0.011 & 11.14$\pm$0.45&  - &  -\\
& & - & - & 0.928$\pm$0.018 & 8.83$\pm$1.02\\ 
& & - & - & 1.633$\pm$0.233 & 21.91$\pm$1.10\\ \hline
55466.144 &  & - & -&   - &  - \\
& & - & - & 0.910$\pm$0.009 & 8.54$\pm$0.90\\
& & - & - & 1.260$\pm$0.197 & 18.82$\pm$0.93\\ \hline
55466.211 &  & - & -&  - &  -\\
& & - & - & 0.967$\pm$0.015 & 7.17$\pm$0.94\\
& & - & - & 1.632$\pm$0.305 & 22.09$\pm$0.95\\ \hline
55466.346 &  & - & -&  - &  -\\
& & - & - & 0.987$\pm$0.008 & 7.75$\pm$0.86\\
& & - & - & 1.640$\pm$0.243 & 18.52 $\pm$1.00\\ \hline
55466.415 &  & 0.040$\pm$0.015 & 6.08$\pm$0.49 &  - &  -\\
& & 1.040$\pm$0.048 & 5.00$\pm$0.77 & 1.011$\pm$0.014 & 10.20$\pm$0.93\\
& & - & - & 1.29$\pm$0.23 & 18.44$\pm$1.20\\ \hline
55466.485 &  &  0.036$\pm$0.013 & 10.86$\pm$0.81 & - &  -\\
& & 1.120$\pm$0.040 & 5.39$\pm$0.72 & 1.090$\pm$0.015 & 7.45$\pm$1.16\\
& & - & - & 1.63$\pm$0.31 & 20.88$\pm$1.13\\ \hline
55466.552 &  &  - & -& - &  -\\
& & - & - & 1.040$\pm$0.022 & 9.75$\pm$0.98\\
& & - & - & 2.170$\pm$0.330 & 19.55$\pm$1.07\\\hline
55467.297$^*$ & 00434928005 & 0.007$\pm$0.002 & 10.0$\pm$0.80 & 0.014$\pm$0.012 & 7.42$\pm$0.40 \\
& & 0.20$\pm$0.03 & 5.48$\pm$0.46 & 0.85$\pm$0.42 & 11.40$\pm$0.97 \\
& & - & - & 1.810$\pm$0.040 & 8.83$\pm$1.13\\
& & - & - &3.390$\pm$0.101 & 6.00$\pm$0.92\\ \hline
55468.2$^*$ & 00434928007 & - & - & - & -\\
& & 0.210$\pm$0.092 & 5.66$\pm$0.71 & - & -\\
& & - & - & 2.410$\pm$0.060 & 7.07$\pm$1.20\\
& & - & -& - & -\\ \hline
55469.2 & 00434928008 & 0.022$\pm$0.013 & 10.49$\pm$0.95 & - &- \\
 & & - & -& - & -\\
& & - & -& - & -\\ \hline
55470.2 & 00434928009 & - & - & - & -\\
& & - & - & - & -\\
& & - & - & - & -\\ \hline
55471.1  & 00434928010 & 0.031$\pm$0.011 & 10.20$\pm$0.78 & - & -\\
& & - & - & - & -\\
& & - & - & - & -\\ \hline
55472.1 & 00434928011 & 0.024$\pm$0.003 & 6.63$\pm$0.53 & - & -\\
& & - & - & 3.64$\pm$0.14 & 11.18$\pm$1.48\\
& & - & - & - & -\\ \hline
55473.1  & 00434928012 & 0.026$\pm$0.003 & 4.47$\pm$0.34 & - & -\\
& & - & - & - & -\\
& & - & - & - & -\\ \hline

55474.1  & 00434928013 & 0.026$\pm$0.004 & 4.58$\pm$0.44 & - & -\\
& & - & - & - & -\\
& & - & - & - & -\\ \hline

55475.1  & 00434928014 & 0.042$\pm$0.007 & 2.65$\pm$0.38 - & -\\
& & - & - & - & -\\
& & - & - & - & -\\ \hline
    \end{tabular}
\end{table*}

\end{document}